\documentclass[11pt,a4paper]{article}
\usepackage{jheppub}

\usepackage[latin1]{inputenc}
\usepackage[T1]{fontenc}
\usepackage{units}
\usepackage{dcolumn}
\usepackage{float,lscape}
\usepackage{array}
\usepackage{multirow}
\def\be{\begin{equation}}
\def\ee{\end{equation}}

\newcommand{\ba}{\begin{array}{c}}
\newcommand{\baz}{\begin{array}{cc}}
\newcommand{\bad}{\begin{array}{ccc}}
\newcommand{\bav}{\begin{array}{cccc}}
\newcommand{\bea}{\begin{equation} \begin{array}{c}}
\newcommand{\eea}{\end{array} \end{equation}}
\newcommand{\ea}{\end{array}}

\newcommand\varmp{\mathbin{\vcenter{\hbox{\oalign{\hfil$\scriptstyle-$\hfil\cr\noalign{\kern-.3ex} $\scriptscriptstyle({+})$\cr}}}}}

\title{\Large\bf Determination of the $\theta_{23}$ octant in LBNO}
\author[a,b,c]{C.R. Das,}
\author[a]{Jukka Maalampi}
\author[b]{Jo\~{a}o Pulido}
\author[a]{Sampsa Vihonen}
\affiliation[a]{Department of Physics, University of Jyv\"askyl\"a\\
P.O.~Box~35, FIN-40014 Jyv\"askyl\"a, Finland,}
\affiliation[b]{Centro de F\'isica Te\'orica das Part\'iculas (CFTP)
Departamento de F\'isica,\\ Instituto Superior T\'ecnico
Av. Rovisco Pais, P-1049-001 Lisboa, Portugal}
\affiliation[c]{Theoretical Physics Division, Physical Research Laboratory,\\
Navrangpura, Ahmedabad - 380 009, India}
\emailAdd{crdas@cftp.ist.utl.pt, crdas@prl.res.in}
\emailAdd{jukka.maalampi@jyu.fi}
\emailAdd{pulido@cftp.ist.utl.pt}
\emailAdd{sampsa.p.vihonen@student.jyu.fi}
\abstract{According to the recent results of the neutrino oscillation experiment MINOS, the neutrino mixing angle $\theta_{23}$ may not be maximal ($45^{\circ}$). Two nearly degenerate solutions are possible, one in the lower octant (LO) where $\theta_{23}<45^{\circ}$, and one in the higher octant (HO) where $\theta_{23}>45^{\circ}$. Long baseline experiments measuring the $\nu_{\mu}\rightarrow\nu_{e}$ are capable of resolving this degeneracy. In this work we study the potential of the planned European LBNO experiment to distinguish between the LO and HO solutions.}
\arxivnumber{1411.2829}
\begin{document}
\maketitle


\section{Introduction}
\label{Introduction}

Neutrino oscillations are described in terms of six physical variables: the three mixing angles $\theta_{12}$, $\theta_{23}$ and $\theta_{13}$, the Dirac CP phase $\delta_{\rm CP}$, and the two squared-mass differences $\Delta m_{21}^2=m_2^2-m_1^2$ and $\Delta m_{31}^2=m_3^2-m_1^2$. Most of these quantities are experimentally measured with good accuracy \cite{Forero, Fogli1, Gonzalez}. However, one still does not know the sign of $\Delta m_{31}^2$, that is, whether the masses obey the normal hierarchy (NH) $m_{1}, m_{2}<m_{3}$ or the inverted hierarchy (IH) $m_{1}, m_{2}>m_{3}$ , and the value of the CP phase $\delta_{\rm CP}$ also is undetermined, all values in the range $-180^{\circ}$ to $+180^{\circ}$ being still allowed. In addition of these two unknowns, which are expected to be resolved in the future long baseline oscillation experiments, there is a third intriguing question known as the octant or $\theta_{23}$-degeneracy problem \cite{Fogli2, raut1, raut2, raut3}. In the leading order the muon neutrino disappearance in the transition $\nu_{\mu}\rightarrow\nu_{\mu}$ is not sensitive to the octancy of $\theta_{23}$, that is, whether this angle lies in the lower octant (LO) $\theta_{23}<45^{\circ}$ or in the higher octant (HO) $\theta_{23}>45^{\circ}$, both alternatives giving the same disappearance probability. In contrast, the leading term of the probability of the electron neutrino appearance $\nu_{\mu}\rightarrow\nu_{e}$ is octant sensitive \cite{Bora:2014zwa}. Hence an accurate measurement of the transition probability $P(\nu_{\mu}\rightarrow\nu_{e})$ in the future long baseline neutrino oscillation experiments might be capable of resolving the octant degeneracy.

Of course, the octant degeneracy problem would not exist if the angle $\theta_{23}$ mixing were maximal, i.e. $\theta_{23}=45^{\circ}$. The recent results of the MINOS oscillation experiment \cite{MINOS} seem to indicate, however, that this is not the case. Two degenerate solutions were found, one in the lower octant (LO) with $\sin^2\theta_{23}\simeq 0.4$ and one in the higher octant (HO) with $\sin^2\theta_{23}\simeq 0.6$. This corresponds to a deviation of about $5^{\circ}$ downwards or upwards, correspondingly, from the maximal-mixing value $\theta_{23}=45^{\circ}$. On the other hand the T2K collaboration \cite{T2K} has recently reported a $\theta_{23}$ central value lying close to the borderline between both octants, being unable to exclude any of the two possibilities.

The octant affects the event rates both for the neutrino transition $\nu_{\mu}\rightarrow\nu_{e}$ and the antineutrino transition $\bar\nu_{\mu}\rightarrow\bar\nu_{e}$, the HO corresponding to more events than the LO. This is true for both the NH and IH mass hierarchies. In addition to the dependence on the $\theta_{23}$ octant, the event rates are quite strongly affected by the value of the CP phase $\delta_{\rm CP}$. A balanced neutrino and anti-neutrino data is requisite for the separation of LO and HO (for a recent analysis, see \cite{Agarwalla}).

In this paper, we analyze the potential of the planned long baseline neutrino oscillation experiment LBNO \cite{Stahl} for resolving the $\theta_{23}$ octant degeneracy. In LBNO the aim is to send neutrino and antineutrino beams, produced at the CERN SPS accelerator, towards the Pyh\"asalmi mine, located in central Finland at the distance 2288 km from CERN, where they will be measured using a two-phase Liquid Argon Time Projection Chamber (LArTPC) \cite{Rubbia1, Rubbia2} combined with a magnetized muon detector (MIND) \cite{Abe, Cervera}. In the first phase, the size of the LArTPC detector is planned to have 20 kton fiducial mass. In this phase a 0.75 MW conventional neutrino beam from the CERN SPS will be used. In the second phase the total detector mass will be extended to 70 kton, and a powerful 2 MW HPPS proton driver \cite{Papaphilippou} is foreseen to be in use. We will determine for both phases the 1$\,\sigma$, 2$\,\sigma$ and 3$\,\sigma$ sensitivity limit of the angle $\theta_{23}$ that LBNO can achieve with 5+5 -years neutrino and antineutrino run, allowing the CP phase to vary in the range $-180^{\circ}$ to $+180^{\circ}$.

\section{Numerical analysis}
\label{Numerical analysis}

The sensitivity for determining the $\theta_{23}$ octant has been previously analysed for NO$\nu$A \cite{NOvA} and T2K \cite{T2K}, as well as for the proposed very long baseline experiment of the future, LBNO \cite{LBNO} and LBNE \cite{Bora:2014zwa}. According to recent reviews (see e.g. \cite{Agarwalla}), the LBNO offers the best potential for determining the octancy of $\theta_{23}$. In this work we present a detailed numerical analysis for the LBNO. The numerical simulation method we use is in most parts adopted from \cite{LBNO}, however, using for our calculations the GLoBES simulation program \cite{GLOBES1,GLOBES2} instead of Monte Carlo simulations.

GLoBES is a simulator that predicts the propagation of neutrinos from the moment they are created in the source to the point they reach the detector and interact with its content. The software evaluates the effect of matter potentials induced by the traversed medium and calculates the resulting event rates that follow from the detection and reconstruction of neutrino events that take place in the detector. The estimated event rates are then used to evaluate the likelihood of different oscillation parameter values with $\chi^2$ -distributions.

The muon neutrino beam is assumed to be produced in the CERN SPS accelerator with a power of 750 kW, shared between neutrino and antineutrino modes at a 50\%/50\% ratio. (This is the same set-up proposed in \cite{LBNO} for the determination of the mass hierarchy.) This corresponds to $1.125\times10^{20}$ POT per year for each beam, and it builds up a total yield of $1.125\times10^{21}$ over the course of the 5+5 -year running time. We also consider the HPPS setup by increasing the annual POT number to $3.0\times10^{21}$. The key parameters concerning the LBNO are presented in table \ref{Experiment parameters}.

The muon neutrino beam is assumed to be nearly pure, though it is contaminated by small numbers of electron neutrinos and antineutrinos. The contamination is an irreducible side product of the muon neutrino creation through meson decays in a hadronic beam and it cannot be removed. We have obtained the respective neutrino fluxes from dedicated flux files based on a GEANT4 simulation \cite{FLUX}.

\begin{table}[h!]
\begin{center}
\begin{tabular}{|c|c|}
\hline
Parameter&Value\\ \hline
Beam power [SPS] ($10^{20}$ POT/yr)&1.125\\
Beam power [HPPS] ($10^{21}$ POT/yr)&3.0\\
Baseline length (km)&2288\\
Running times (yr)&5+5\\
Detection efficiency (\%)&90\\
$\nu_\mu$ NC rejection (\%)&99.5\\
$\nu_\mu$ CC rejection (\%)&99.5\\
Energy resolution (GeV)&$0.15\times\sqrt{E}$\\
Energy window (GeV)&$[0.1,\,10.0]$\\
Number of bins&80\\
Bin width (GeV)&0.125\\ \hline
\end{tabular}
\end{center}
\caption{Experiment parameters}
\label{Experiment parameters}
\end{table}

In this work we assume the following detecting properties \cite{Li1,Li2}. The LArTPC detector is capable of detecting electron and muon neutrinos by observing secondary electron and muon leptons at approximately constant 90\% rate. The LBNO neutrinos are detected within [0.1 GeV, 10.0 GeV] energy range, which is divided into 80 energy bins, each bin 0.125 GeV wide. The detection and reconstruction process has the following parameters: Whenever a neutrino interacts with the detector substance, the counting system reconstructs the energy and flavor of the incident neutrino and identifies the event with the corresponding energy bin. The reconstructed energy is assumed to be normally distributed with a resolution of $0.15\times\sqrt{E}$, where \textit{E} is the neutrino energy in GeV. The cross sections of the charged current (CC) and neutral current (NC) neutrino-nucleon interactions are given in cross section files simulated for LArTPC with a dedicated GENIE simulation \cite{X-SEC}. The simulation is specifically dedicated to LArTPC systems, and it takes the oscillations to tau neutrinos into account better than any previous simulation.

The LBNO experiment is designed to study the electron appearance probabilities $P(\nu_\mu \rightarrow \nu_e)$ and $P(\bar{\nu}_\mu \rightarrow \bar{\nu}_e)$ by counting the corresponding CC events in the detector. These CC events constitute the signal, whereas background consists of any type of events that have similar final state properties. On one hand, the electron appearance channels gain background from CC and NC events with $\nu_e$ and $\bar{\nu}_e$ arising from the oscillations of the impurities in the muon neutrino beam. On the other hand, the detector is also assumed to have a $0.5\%$ chance to accept $\nu_\mu$ and $\bar{\nu}_\mu$ from $\nu_\mu\rightarrow\nu_\mu$ and $\bar\nu_\mu\rightarrow\bar\nu_\mu$ as $\nu_e$ and $\bar{\nu}_e$ from both CC and NC event categories. Lastly, $\nu_\tau$ and $\bar{\nu}_\tau$ neutrinos originated from $\nu_\mu \rightarrow \nu_\tau$ and $\bar{\nu}_\mu \rightarrow \bar{\nu}_\tau$ oscillations also contribute to the background. The number of $\tau$ leptons produced in the detector from these neutrinos accounts for approximately 6\% of the total number of leptons produced \cite{Stahl}. Inserting the branching ratio of the $\tau$ subsequent decay into electrons through $\tau \rightarrow e\,\nu_e\,\nu_\tau$ ($\sim$17.8\%) \cite{PDG} together with the detector efficiency (90\%), one sees that the corresponding $\nu_e$ contamination is 1\%, hence of the same order of magnitude as the intrinsic beam contamination.

Besides electron appearance, also the muon disappearance probabilities $P(\nu_\mu \rightarrow \nu_\mu)$ and $P(\bar{\nu}_\mu \rightarrow \bar{\nu}_\mu)$ are studied. In this case the signal composes of the respective CC events whereas the background consists of $\nu_\mu \rightarrow \nu_\mu$ and $\bar{\nu}_\mu \rightarrow \bar{\nu}_\mu$ NC neutrinos and $\nu_e \rightarrow \nu_\mu$ and $\bar{\nu}_e \rightarrow \bar{\nu}_\mu$ CC neutrinos that are mistakenly accepted as signal neutrinos. We assume in our analysis the experiment to be able to distinguish between neutrinos and antineutrinos, reducing the background of $\nu_\mu \rightarrow \nu_\mu$ and $\nu_\mu \rightarrow \nu_e$ to only neutrino channels and $\bar{\nu}_\mu \rightarrow \bar{\nu}_\mu$ and $\bar{\nu}_\mu \rightarrow \bar{\nu}_e$ only to antineutrino channels, respectively. Also, $\nu_\mu\rightarrow\nu_\tau$ and $\bar\nu_\mu\rightarrow\bar\nu_\tau$ oscillations contribute to $\nu_\mu$ and $\bar\nu_\mu$ background through $\tau\rightarrow\mu\,\nu_\mu\,\nu_\tau$ decay.

The $\chi^2$ values are calculated as follows (see e.g. \cite{GLOBES1,GLOBES2}). The statistical part is computed with the Poissonian function
\begin{equation}
\chi^2 (\omega,\omega_0) = \sum_{i=1}^{80}{2\left[T_i-O_i\left(1-\textrm{ln}\frac{T_i}{O_i}\right)\right]},
\label{CHI001}
\end{equation}
where the number of observed events ($O_i$) in the $i\textrm{th}$ bin is computed from the so called true values ($\omega_0$) and the number of test events ($T_i$) from the test values ($\omega$), respectively.

The observed events is the category of events that would result from oscillation parameter values that one considers to be closest to the truth. They are based on the best-fit values obtained from the most recent experiments. We denote these values with $\omega_0$. Since all parameter values are not precisely known, such as the sign of $\Delta m_{31}^2$, the $\chi^2$ values need to be computed for all possible scenarios. The number of observed events is taken to be the sum of events from signal and background components:
\begin{equation}
O_i = N_i^\textrm{sg} (\omega_0) + N_i^\textrm{bg} (\omega_0),
\label{EVT001}
\end{equation}
where $N_i^\textrm{sg}$ and $N_i^\textrm{bg}$ stand for the numbers of signal and background events.

The test values on the other hand stand for event numbers that are computed with whatever oscillation parameter values one wants to test. We denote these values with $\omega$. We also apply systematic errors to both signal and background events by incorporating two nuisance parameters \cite{GLOBES1,GLOBES2}, $\zeta_1$ and $\zeta_2$, with error weights $\pi_1$ and $\pi_2$:
\begin{equation}
T_i = N_i^\textrm{sg} (\omega) [1+\pi_1 \zeta_1] + N_i^\textrm{bg} (\omega) [1+\pi_2 \zeta_2].
\label{EVT002}
\end{equation}

The systematic errors are included by minimizing the $\chi^2$ function over nuisance parameters $\zeta_1$ and $\zeta_2$:
\begin{equation}
\chi_\textrm{pull}^2 (\omega,\omega_0) = \min_{\zeta_1,\zeta_2} \left[\chi^2 (\omega,\omega_0)+\zeta_1^2+\zeta_2^2\right],
\label{CHI002}
\end{equation}
where $\chi^2 (\omega,\omega_0)$ is the Poissonian function given by equation (\ref{CHI001}). We assume $5\,\%$ systematical error weights in both signal and background by setting $\pi_1, \pi_2 = 0.05$. This corresponds to the normalization error in the LArTPC detectors \cite{Agarwalla}.

We also assume that the values of $\theta_{12}$, $\theta_{13}$, $\Delta m_{21}^2$, $\Delta m_{31}^2$, $\delta_\textrm{CP}$ and $\rho$ are associated with standard deviations $\sigma(\theta_{12})$, $\sigma(\theta_{13})$, $\sigma(\Delta m_{21}^2)$, $\sigma(\Delta m_{31}^2)$ and $\sigma(\rho)$. We include these parameter uncertainties via the so called priors \cite{GLOBES1,GLOBES2}. The prior function is given by:
\begin{eqnarray}
\chi_\textrm{prior}^2 (\omega,\omega_0) & = & \left(\frac{\sin^2 \theta_{12} (\omega)-\sin^2 \theta_{12} (\omega_0)}{\sigma(\sin^2 \theta_{12})}\right)^2 + \left(\frac{\sin^2 2\theta_{13} (\omega)-\sin^2 2\theta_{13} (\omega_0)}{\sigma(\sin^2 2\theta_{13})}\right)^2\nonumber\\ && + \left(\frac{\Delta m_{21}^2 (\omega)-\Delta m_{21}^2 (\omega_0)}{\sigma(\Delta m_{21}^2)}\right)^2 + \left(\frac{\Delta m_{31}^2 (\omega)-\Delta m_{31}^2 (\omega_0)}{\sigma(\Delta m_{31}^2)}\right)^2\nonumber\\ && + \left(\frac{\rho (\omega)-\rho(\omega_0)}{\sigma(\rho)}\right)^2.
\label{CHI003}
\end{eqnarray}

\begin{table}[ht!]
\begin{center}
\begin{tabular}{|c|c|c|}
\hline
Parameter&Value&Error (1$\,\sigma$)\\ \hline
$\sin^2 \theta_{12}$&0.304&0.013\\
$\sin^2 \theta_{13}$ (NH)&0.0218&0.0010\\
$\sin^2 \theta_{13}$ (IH)&0.0219&0.0011\\
$\sin^2 \theta_{23}$&varied&0\\
$\delta_\textrm{CP}$ [$^\circ$]&varied&0\\
$\Delta m_{21}^2$ [$10^{-5}\,\textrm{eV}^2$]&7.50&0.19\\
$\Delta m_{31}^2$ (NH) [$10^{-3}\,\textrm{eV}^2$]&2.457&0.047\\
$\Delta m_{31}^2$ (IH) [$10^{-3}\,\textrm{eV}^2$]&-2.449&0.048\\ \hline
\end{tabular}
\end{center}
\caption{Oscillation parameters}
\label{Oscillation parameters}
\end{table}

\noindent The overall $\chi^2$ value is calculated as the sum of the pull and prior parts from equations (\ref{CHI002}) and (\ref{CHI003}), which is then minimized over the test values:
\begin{equation}
\chi_\textrm{total}^2 (\omega_0) = \min_\omega \left[\chi_\textrm{pull}^2 (\omega,\omega_0) + \chi_\textrm{prior}^2 (\omega,\omega_0)\right].
\label{CHI004}
\end{equation}

The matter density parameter $\rho$ is taken into account as a variable in equation (\ref{CHI004}). The density distribution of the Earth's crust between CERN and Pyh\"asalmi is known to a good accuracy \cite{Kozlovskaya}, but for this study we consider it sufficient to evaluate the matter density function with a 20-step PREM distribution \cite{PREM}, and assume 2\% error value (1$\,\sigma$).

The final $\chi^2$ value is calculated by minimizing $\chi_\textrm{total}^2$ over all oscillation parameters in the test values, that is, over $\omega$. The prior function constrains the value ranges over which $\chi_\textrm{total}^2$ may converge, and the absence of $\delta_\textrm{CP}$ in equation (\ref{CHI003}) indicates that no such constraints are assumed for $\delta_\textrm{CP}$. We also choose to keep $\theta_{23}$ fixed in the minimization process.

We calculate the 1$\,\sigma$, 2$\,\sigma$ and 3$\,\sigma$ confidence levels for the event that the LBNO experiment will be able to rule out one octant when the other octant is assumed to be correct. This is done by computing $\chi^2$ first for $\theta_{23}$ and then 90$^\circ - \theta_{23}$, and calculating the difference between the two $\chi^2$ values, both calculated as given in equation (\ref{CHI004}):
\begin{equation}
\Delta \chi^2 = \chi_\textrm{total}^2 (90^\circ-\theta_{23})-\chi_\textrm{total}^2 (\theta_{23}).
\label{CHI005}
\end{equation}

\noindent We take the true values from \cite{PARAMS}, which contains a recent compilation on experimentally determined parameter values. These values are also presented in table \ref{Oscillation parameters}.

The Gaussian errors shown in table \ref{Oscillation parameters} are distributed for parameters $\sin^2 \theta_{12}$, $\sin^2 2\theta_{13}$, $\Delta m_{21}^2$ and $\Delta m_{31}^2$, respectively. The errors of $\delta_\textrm{CP}$ and $\sin^2 \theta_{23}$ are not present in the prior function $\chi_\textrm{prior}^2$ and therefore they are both marked with zero. This follows from our choice that $\delta_\textrm{CP}$ is not assigned with constraints and $\theta_{23}$ is kept fixed in the minimization of $\chi^2$.

The minimization of the $\chi^2$ function in equation (\ref{CHI004}) is carried out keeping $\theta_{23}$ fixed and other parameters free. Since $\theta_{23}$ and $\delta_\textrm{CP}$ are not precisely known, we calculate the $\chi^2$ values for different possible values of $\theta_{23}$ and $\delta_\textrm{CP}$, and for both mass hierarchies as well.

\section{Results and discussion}
\label{Results and discussion}
We have investigated the ability of the LBNO experiment to determine the octancy of the neutrino mixing angle $\theta_{23}$ up to a 3$\,\sigma$ confidence limit (CL) for all values of the phase $\delta_\textrm{CP}$. This was done by computing the $\Delta \chi^ 2$ distribution for a range of $\theta_{23}$ and $\delta_\textrm{CP}$ values. The $\Delta \chi^2$ distribution was computed with a grid of 120$\times$360 points, interpolating the intermediate values.

The contour plots were produced for four different setups: SPS beam with 20 kt LArTPC, SPS beam with 70 kt LArTPC, HPPS beam with 20 kt LArTPC and HPPS beam with 70 kt LArTPC. Figures 1 and 2 present the resulting 1$\,\sigma$, 2$\,\sigma$ and 3$\,\sigma$ CL contours for the normal and inverted hierarchy, respectively. In each figure the white regions in the plots are the areas for which the values of $\theta_{23}$, $\delta_\textrm{CP}$ can be established with CL greater than 3$\,\sigma$. So for all $\theta_{23}$, $\delta_\textrm{CP}$ data points in these areas, one can eliminate with a CL larger than 3$\,\sigma$ the possibility for these parameters to lie in the other octant. Conversely the coloured regions illustrate the cases where no such distinction is possible with the indicated CL. Some details of these contours are presented numerically in table \ref{Results3}.

\begin{figure}[ht!]
\begin{center}
\includegraphics[width=\textwidth]{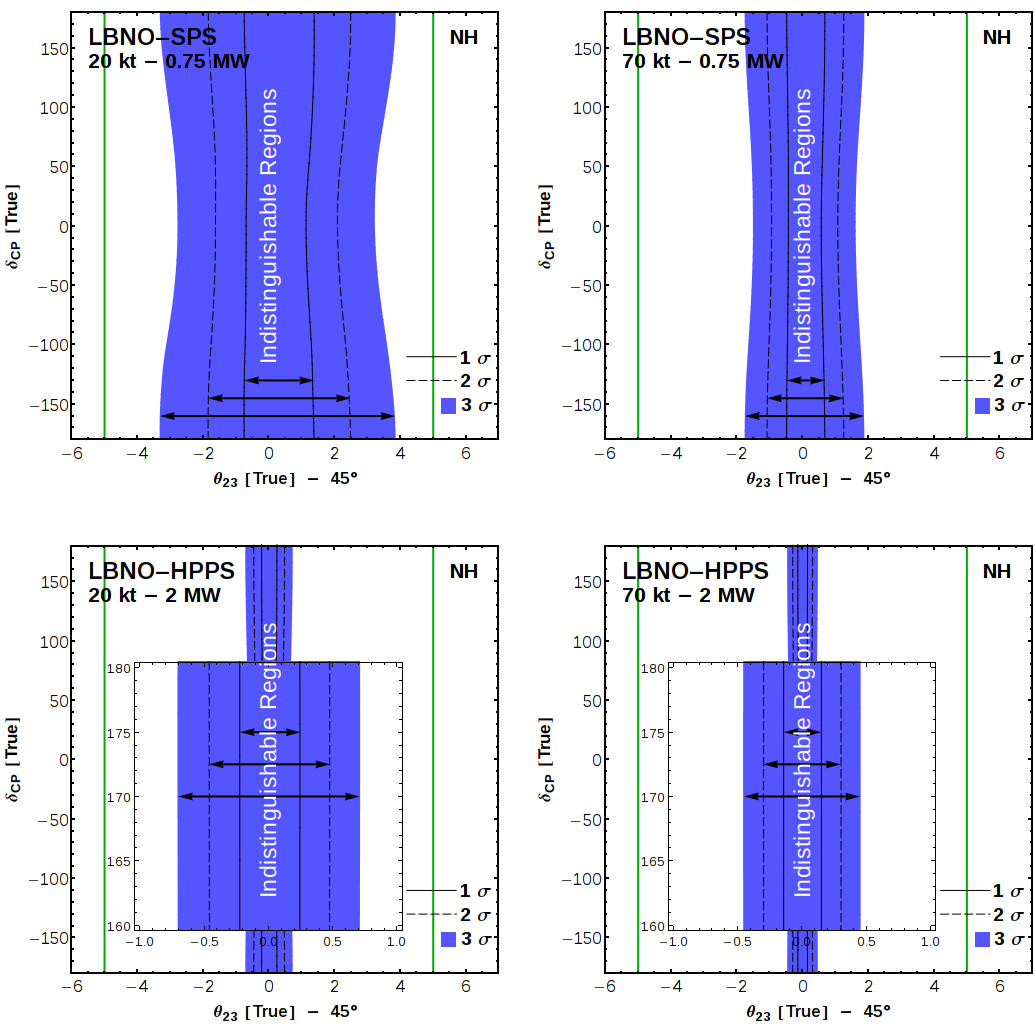}
\end{center}
\caption{Octant discovery potential in the LBNO when normal mass hierarchy (NH) is assumed. The coloured regions show values of $\theta_{23}$ and $\delta_\textrm{CP}$ where the two octant solutions can be distinguished from each other in the LBNO at less than 3$\,\sigma$ confidence level (i.e. the two octants are indistinguishable up to this limit). The white areas show the values where the other octant can be rejected at 3$\,\sigma$ or better. The 1$\,\sigma$ and 2$\,\sigma$ contours are shown with solid and dashed lines, respectively. The MINOS favoured $\theta_{23} - 45^\circ = \pm 5^\circ$ values are marked with green lines.}
\label{Results1}
\end{figure}

\begin{figure}[ht!]
\begin{center}
\includegraphics[width=\textwidth]{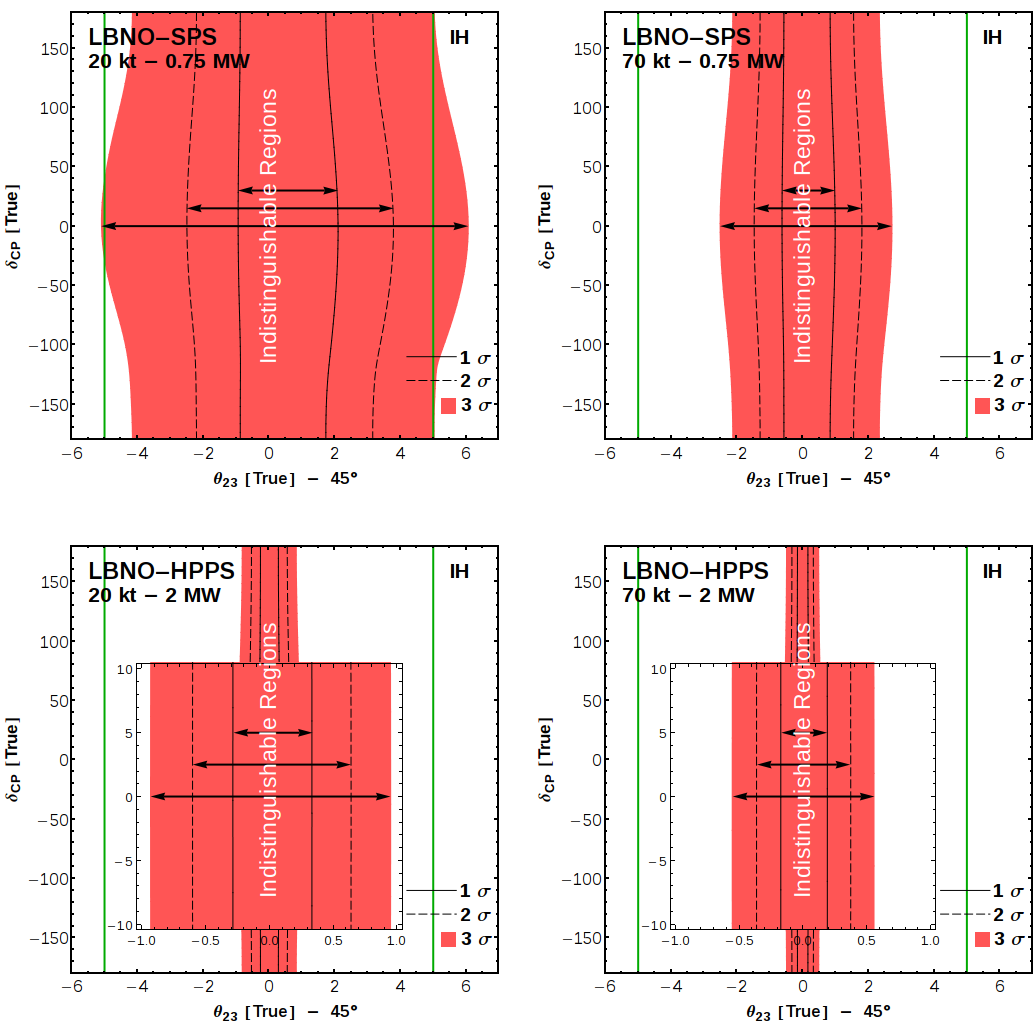}
\end{center}
\caption{Octant discovery potential in the LBNO when inverted mass hierarchy (IH) is assumed. The coloured regions show values of $\theta_{23}$ and $\delta_\textrm{CP}$ where the two octant solutions can be distinguished from each other in the LBNO at less than 3$\,\sigma$ confidence level (i.e. the two octants are indistinguishable up to this limit). The white areas show the values where the other octant can be rejected at 3$\,\sigma$ or better. The 1$\,\sigma$ and 2$\,\sigma$ contours are shown with solid and dashed lines, respectively. The MINOS favoured $\theta_{23} - 45^\circ = \pm 5^\circ$ values are marked with green lines.}
\label{Results2}
\end{figure}

We have marked in figures \ref{Results1} and \ref{Results2} by green lines the MINOS favoured $\theta_{23}$ values 40$^{\circ}$ and 50$^{\circ}$. It is seen that for all the different setups considered the right $\theta_{23}$ octant can be asserted in NH with at least 3$\,\sigma$ CL. As for IH this limit is reached for the lower octant in all cases, whereas for the higher octant it fails to be reached in the sole case of the 20 kt setup with 0.75 MW. All other setup versions yield improved sensitivities so that the 3$\,\sigma$ limit can be reached for all of them regardless of the mass hierarchy and $\delta_\textrm{CP}$ value. The graphs also show by themselves that increasing beam power (by a factor of 2.7 i.e. from 0.75 MW to 2 MW) with the same detector is a lot more effective than increasing detector size from 20kt to 70 kt with the same beam power.

We also studied the scenario in which the neutrino and antineutrino beam modes are divided by a 75\%/25\% ratio, which has been suggested to optimize the LBNO for the CP violation search \cite{LBNO}. Our results show decreased sensitivity for determining the $\theta_{23}$ octant. Furthermore we also found that a 25\%/75\% share between neutrinos and antineutrinos improves the sensitivity to $\theta_{23}$ octant determination relative to the 50\%/50\% share. Hence the shorter running times with neutrinos combined with the longer running times with antineutrinos is found to improve the octant sensitivity, whereas the opposite combination worsens it.

In principle, any increase in the exposure moves the 3$\,\sigma$ CL contour closer to the $\theta_{23} = 45^{\circ}$ value. If one is to expect that the real value of $\theta_{23}$ is to be 5$^{\circ}$ off from 45$^{\circ}$, then even the SPS setup with 20 kt detector may be sufficient to reach the 3$\,\sigma$ CL for both mass hierarchies. The sensitivity is worse near 45$^{\circ}$, however, and it would require an upgrade to reach the 3$\,\sigma$ CL benchmark. A future HPPS facility with a 70 kt LArTPC detector, for instance, could solve this problem as it would set the limit to less than $\pm0.6^{\circ}$. Furthermore, LBNO will most likely be able to measure the mass hierarchy with a 0.75 MW SPS beam and a 20 kt LArTPC detector, in which case the acquired data could be used to narrow down the estimate on the $\theta_{23}$ octant. The determination of the $\theta_{23}$ octancy would hence be a logical follow-up of the mass hierarchy measurement.

\newpage

\newpage
\begin{landscape}
\begin{table}[p]
\centering
{\scriptsize
\begin{tabular}{|c|c|c|c|c|c|c|c|c|c|c|c|c|c|c|c|c|}
\multicolumn{17}{c}{}\\
\hline
\multicolumn{1}{|c|}{}&\multicolumn{8}{c}{}&\multicolumn{8}{|c|}{}\\
\parbox[t]{3mm}{\multirow{3}{*}{\rotatebox[origin=c]{90}{\scriptsize\bf Indistinguishable Regions}}}&\multicolumn{8}{c}{\bf 20 kt Detector}&\multicolumn{8}{|c|}{\bf 70 kt Detector}\\
\multicolumn{1}{|c|}{}&\multicolumn{8}{c}{}&\multicolumn{8}{|c|}{}\\
\cline{2-17}

\multicolumn{1}{|c|}{}&\multicolumn{4}{c}{}&\multicolumn{4}{|c}{}&\multicolumn{4}{|c}{}&\multicolumn{4}{|c|}{}\\
\multicolumn{1}{|c|}{}&\multicolumn{4}{c}{\bf SPS (0.75 MW)}&\multicolumn{4}{|c}{\bf HPPS (2 MW)}&\multicolumn{4}{|c}{\bf SPS (0.75 MW)}&\multicolumn{4}{|c|}{\bf HPPS (2 MW)}\\
\multicolumn{1}{|c|}{}&\multicolumn{4}{c}{}&\multicolumn{4}{|c}{}&\multicolumn{4}{|c}{}&\multicolumn{4}{|c|}{}\\
\cline{2-17}

\multicolumn{1}{|c|}{}&\multicolumn{2}{c}{}&\multicolumn{2}{|c}{}&\multicolumn{2}{|c}{}&\multicolumn{2}{|c}{}&\multicolumn{2}{|c}{}&\multicolumn{2}{|c}{}&\multicolumn{2}{|c}{}&\multicolumn{2}{|c|}{}\\
\multicolumn{1}{|c|}{}&\multicolumn{2}{c}{\bf NH}&\multicolumn{2}{|c}{\bf IH}&\multicolumn{2}{|c}{\bf NH}&\multicolumn{2}{|c}{\bf IH}&\multicolumn{2}{|c}{\bf NH}&\multicolumn{2}{|c}{\bf IH}&\multicolumn{2}{|c}{\bf NH}&\multicolumn{2}{|c|}{\bf IH}\\
\multicolumn{1}{|c|}{}&\multicolumn{2}{c}{}&\multicolumn{2}{|c}{}&\multicolumn{2}{|c}{}&\multicolumn{2}{|c}{}&\multicolumn{2}{|c}{}&\multicolumn{2}{|c}{}&\multicolumn{2}{|c}{}&\multicolumn{2}{|c|}{}\\
\multicolumn{1}{|c|}{}&\multicolumn{2}{c}{$\theta_{23}\,{\rm [True]} - 45^\circ$}&\multicolumn{2}{|c}{$\theta_{23}\,{\rm [True]} - 45^\circ$}&\multicolumn{2}{|c}{$\theta_{23}\,{\rm [True]} - 45^\circ$}&\multicolumn{2}{|c}{$\theta_{23}\,{\rm [True]} - 45^\circ$}&\multicolumn{2}{|c}{$\theta_{23}\,{\rm [True]} - 45^\circ$}&\multicolumn{2}{|c}{$\theta_{23}\,{\rm [True]} - 45^\circ$}&\multicolumn{2}{|c}{$\theta_{23}\,{\rm [True]} - 45^\circ$}&\multicolumn{2}{|c|}{$\theta_{23}\,{\rm [True]} - 45^\circ$}\\
\multicolumn{1}{|c|}{}&\multicolumn{2}{c}{}&\multicolumn{2}{|c}{}&\multicolumn{2}{|c}{}&\multicolumn{2}{|c}{}&\multicolumn{2}{|c}{}&\multicolumn{2}{|c}{}&\multicolumn{2}{|c}{}&\multicolumn{2}{|c|}{}\\
\multicolumn{1}{|c|}{}&
\multicolumn{1}{c}{$(-)$}&\multicolumn{1}{c}{$(+)$}&\multicolumn{1}{|c}{$(-)$}&\multicolumn{1}{c}{$(+)$}&\multicolumn{1}{|c}{$(-)$}&\multicolumn{1}{c}{$(+)$}&\multicolumn{1}{|c}{$(-)$}&\multicolumn{1}{c}{$(+)$}&\multicolumn{1}{|c}{$(-)$}&\multicolumn{1}{c}{$(+)$}&\multicolumn{1}{|c}{$(-)$}&\multicolumn{1}{c}{$(+)$}&\multicolumn{1}{|c}{$(-)$}&\multicolumn{1}{c}{$(+)$}&\multicolumn{1}{|c}{$(-)$}&\multicolumn{1}{c|}{$(+)$}\\
\multicolumn{1}{|c|}{}&\multicolumn{2}{c}{}&\multicolumn{2}{|c}{}&\multicolumn{2}{|c}{}&\multicolumn{2}{|c}{}&\multicolumn{2}{|c}{}&\multicolumn{2}{|c}{}&\multicolumn{2}{|c}{}&\multicolumn{2}{|c|}{}\\
\hline

\multicolumn{1}{|c|}{}&\multicolumn{1}{c}{}&\multicolumn{1}{|c}{}&\multicolumn{1}{|c}{}&\multicolumn{1}{|c}{}&\multicolumn{1}{|c}{}&\multicolumn{1}{|c}{}&\multicolumn{1}{|c}{}&\multicolumn{1}{|c}{}&\multicolumn{1}{|c}{}&\multicolumn{1}{|c}{}&\multicolumn{1}{|c}{}&\multicolumn{1}{|c}{}&\multicolumn{1}{|c}{}&\multicolumn{1}{|c}{}&\multicolumn{1}{|c}{}&\multicolumn{1}{|c|}{}\\
\multicolumn{1}{|c|}{1$\,\sigma$}&\multicolumn{1}{c}{0.75$^\circ$}&\multicolumn{1}{|c}{1.38$^\circ$}&\multicolumn{1}{|c}{0.94$^\circ$}&\multicolumn{1}{|c}{2.11$^\circ$}&\multicolumn{1}{|c}{0.22$^\circ$}&\multicolumn{1}{|c}{0.25$^\circ$}&\multicolumn{1}{|c}{0.28$^\circ$}&\multicolumn{1}{|c}{0.34$^\circ$}&\multicolumn{1}{|c}{0.48$^\circ$}&\multicolumn{1}{|c}{0.68$^\circ$}&\multicolumn{1}{|c}{0.63$^\circ$}&\multicolumn{1}{|c}{0.99$^\circ$}&\multicolumn{1}{|c}{0.14$^\circ$}&\multicolumn{1}{|c}{0.15$^\circ$}&\multicolumn{1}{|c}{0.17$^\circ$}&\multicolumn{1}{|c|}{0.19$^\circ$}\\
\multicolumn{1}{|c|}{}&\multicolumn{1}{c}{}&\multicolumn{1}{|c}{}&\multicolumn{1}{|c}{}&\multicolumn{1}{|c}{}&\multicolumn{1}{|c}{}&\multicolumn{1}{|c}{}&\multicolumn{1}{|c}{}&\multicolumn{1}{|c}{}&\multicolumn{1}{|c}{}&\multicolumn{1}{|c}{}&\multicolumn{1}{|c}{}&\multicolumn{1}{|c}{}&\multicolumn{1}{|c}{}&\multicolumn{1}{|c}{}&\multicolumn{1}{|c}{}&\multicolumn{1}{|c|}{}\\
\hline

\multicolumn{1}{|c|}{}&\multicolumn{1}{c}{}&\multicolumn{1}{|c}{}&\multicolumn{1}{|c}{}&\multicolumn{1}{|c}{}&\multicolumn{1}{|c}{}&\multicolumn{1}{|c}{}&\multicolumn{1}{|c}{}&\multicolumn{1}{|c}{}&\multicolumn{1}{|c}{}&\multicolumn{1}{|c}{}&\multicolumn{1}{|c}{}&\multicolumn{1}{|c}{}&\multicolumn{1}{|c}{}&\multicolumn{1}{|c}{}&\multicolumn{1}{|c}{}&\multicolumn{1}{|c|}{}\\
\multicolumn{1}{|c|}{2$\,\sigma$}&\multicolumn{1}{c}{1.85$^\circ$}&\multicolumn{1}{|c}{2.49$^\circ$}&\multicolumn{1}{|c}{2.50$^\circ$}&\multicolumn{1}{|c}{3.80$^\circ$}&\multicolumn{1}{|c}{0.46$^\circ$}&\multicolumn{1}{|c}{0.48$^\circ$}&\multicolumn{1}{|c}{0.60$^\circ$}&\multicolumn{1}{|c}{0.64$^\circ$}&\multicolumn{1}{|c}{1.08$^\circ$}&\multicolumn{1}{|c}{1.25$^\circ$}&\multicolumn{1}{|c}{1.47$^\circ$}&\multicolumn{1}{|c}{1.81$^\circ$}&\multicolumn{1}{|c}{0.30$^\circ$}&\multicolumn{1}{|c}{0.30$^\circ$}&\multicolumn{1}{|c}{0.36$^\circ$}&\multicolumn{1}{|c|}{0.37$^\circ$}\\
\multicolumn{1}{|c|}{}&\multicolumn{1}{c}{}&\multicolumn{1}{|c}{}&\multicolumn{1}{|c}{}&\multicolumn{1}{|c}{}&\multicolumn{1}{|c}{}&\multicolumn{1}{|c}{}&\multicolumn{1}{|c}{}&\multicolumn{1}{|c}{}&\multicolumn{1}{|c}{}&\multicolumn{1}{|c}{}&\multicolumn{1}{|c}{}&\multicolumn{1}{|c}{}&\multicolumn{1}{|c}{}&\multicolumn{1}{|c}{}&\multicolumn{1}{|c}{}&\multicolumn{1}{|c|}{}\\
\hline

\multicolumn{1}{|c|}{}&\multicolumn{1}{c}{}&\multicolumn{1}{|c}{}&\multicolumn{1}{|c}{}&\multicolumn{1}{|c}{}&\multicolumn{1}{|c}{}&\multicolumn{1}{|c}{}&\multicolumn{1}{|c}{}&\multicolumn{1}{|c}{}&\multicolumn{1}{|c}{}&\multicolumn{1}{|c}{}&\multicolumn{1}{|c}{}&\multicolumn{1}{|c}{}&\multicolumn{1}{|c}{}&\multicolumn{1}{|c}{}&\multicolumn{1}{|c}{}&\multicolumn{1}{|c|}{}\\
\multicolumn{1}{|c|}{3$\,\sigma$}&\multicolumn{1}{c}{3.31$^\circ$}&\multicolumn{1}{|c}{3.85$^\circ$}&\multicolumn{1}{|c}{5.10$^\circ$}&\multicolumn{1}{|c}{6.07$^\circ$}&\multicolumn{1}{|c}{0.70$^\circ$}&\multicolumn{1}{|c}{0.71$^\circ$}&\multicolumn{1}{|c}{0.93$^\circ$}&\multicolumn{1}{|c}{0.95$^\circ$}&\multicolumn{1}{|c}{1.75$^\circ$}&\multicolumn{1}{|c}{1.87$^\circ$}&\multicolumn{1}{|c}{2.51$^\circ$}&\multicolumn{1}{|c}{2.72$^\circ$}&\multicolumn{1}{|c}{0.45$^\circ$}&\multicolumn{1}{|c}{0.45$^\circ$}&\multicolumn{1}{|c}{0.55$^\circ$}&\multicolumn{1}{|c|}{0.56$^\circ$}\\
\multicolumn{1}{|c|}{}&\multicolumn{1}{c}{}&\multicolumn{1}{|c}{}&\multicolumn{1}{|c}{}&\multicolumn{1}{|c}{}&\multicolumn{1}{|c}{}&\multicolumn{1}{|c}{}&\multicolumn{1}{|c}{}&\multicolumn{1}{|c}{}&\multicolumn{1}{|c}{}&\multicolumn{1}{|c}{}&\multicolumn{1}{|c}{}&\multicolumn{1}{|c}{}&\multicolumn{1}{|c}{}&\multicolumn{1}{|c}{}&\multicolumn{1}{|c}{}&\multicolumn{1}{|c|}{}\\
\hline

\multicolumn{1}{|c|}{}&\multicolumn{2}{c}{}&\multicolumn{2}{|c}{}&\multicolumn{2}{|c}{}&\multicolumn{2}{|c}{}&\multicolumn{2}{|c}{}&\multicolumn{2}{|c}{}&\multicolumn{2}{|c}{}&\multicolumn{2}{|c|}{}\\
\multicolumn{1}{|c|}{\bf Maximum}&\multicolumn{2}{c}{$180^\circ$}&\multicolumn{2}{|c}{$0^\circ$}&\multicolumn{2}{|c}{$180^\circ$}&\multicolumn{2}{|c}{$0^\circ$}&\multicolumn{2}{|c}{$180^\circ$}&\multicolumn{2}{|c}{$0^\circ$}&\multicolumn{2}{|c}{$180^\circ$}&\multicolumn{2}{|c|}{$0^\circ$}\\
\multicolumn{1}{|c|}{\bf at $\delta_{\rm CP}\,{\rm [True]}$}&\multicolumn{2}{c}{}&\multicolumn{2}{|c}{}&\multicolumn{2}{|c}{}&\multicolumn{2}{|c}{}&\multicolumn{2}{|c}{}&\multicolumn{2}{|c}{}&\multicolumn{2}{|c}{}&\multicolumn{2}{|c|}{}\\
\hline
\end{tabular}
}
\caption{Octant discovery potential in LBNO. The highest and lowest $45^{\circ}-\theta_{23}$ values at which the 1$\,\sigma$, 2$\,\sigma$ and 3$\,\sigma$ confidence levels can be reached are shown in the case of normal hierarchy (NH) and inverted hierarchy (IH).}
\label{Results3}
\end{table}
\end{landscape}

\section*{Acknowledgments}
SV would like to thank Kai Loo and Peter Ballet for useful conversations that led to improvements in the GLoBES simulations. CRD sincerely thanks Physical Research Laboratory and Prof. Utpal Sarkar (Dean) for Visiting Scientist position, also greatly thanks the Department of Physics, University of Jyv\"askyl\"a, where the work was initiated, for hospitality and financial support. JP acknowledges financial support from FCT neutrino project PTDC/FIS-NUC/0548/2012 and expresses his thanks to the Department of Physics, University of Jyv\"askyl\"a, for hospitality and partial financial support.


\end{document}